\documentclass[aps,prf,amsfonts,longbibliography,notitlepage]{revtex4-2}
\usepackage{amsmath,amssymb,graphicx,tikz,subeqnarray}
\usepackage{xcolor}
\usepackage[colorlinks=true, citecolor=red, linkcolor=blue, urlcolor=blue ]{hyperref}

\def \bm#1{{\bf #1}}

\def\u{{\bf u}}
\begin{document}
	
	\title{Hydrodynamic interactions between a {point force} and a slender filament}
	
	\author{Ivan Tanasijevi\'c}
	
	\email[]{it279@cam.ac.uk}
	
	\affiliation{Department of Applied Mathematics and Theoretical Physics, Centre for Mathematical Sciences, University of Cambridge, Wilberforce Road, Cambridge CB3 0WA, United Kingdom}
	
	\author{Eric Lauga}
	
	\email[]{e.lauga@damtp.cam.ac.uk}
	
	\affiliation{Department of Applied Mathematics and Theoretical Physics, Centre for Mathematical Sciences, University of Cambridge, Wilberforce Road, Cambridge CB3 0WA, United Kingdom}
	
	\date{\today}
	
	\begin{abstract}
		
		The Green's function of the incompressible Stokes {equations}, the stokeslet,  represents the singular flow due to a point force. Its classical value in an unbounded fluid has been extended near surfaces of various shapes, including flat walls and spheres, 
		and in most cases the presence of a surface leads to an advection flow induced at the location of the point force. In this paper, motivated by the biological  transport of  cargo along polymeric  filaments  inside  eukaryotic cells, 	 we investigate the {reaction flow at the location of the point force     due to a rigid  slender filament located at a separation distance intermediate between the filament radius and its length  (i.e.~we compute the advection of the point force induced by the presence of the filament).} 
		An asymptotic analysis of the problem reveals that the  leading-order approximation for the force distribution along the axis of the filament takes a form analogous to  resistive-force theory but with drag coefficients that  depend logarithmically on the distance between the point force and the filament. A comparison of our theoretical prediction with boundary element computations   show good agreement.  
		We finally briefly  extend the model to the case of curved filaments.
	\end{abstract}

	\maketitle 
	
	\section{Introduction}
	
	As  for many other partial differential equations, the Green's function of the incompressible Stokes equations~\cite{Lorentz} is useful to help solve these equations in a variety of configurations~\cite{KimKarrila,HappelBrenner}. In an unbounded fluid, that Green's function is   refereed to as the stokeslet~\cite{Hancock_stokeslet}. Beyond its fundamental mathematical structure, the stokeslet  represents a fundamental flow; it is  asymptotically the leading-order    flow induced by a forced rigid body when observed sufficiently far from it~\cite{HappelBrenner}. In more complex geometries, one can extend 
	this fundamental solution and its higher multipole moments to satisfy the correct boundary condition on surfaces. Classical work in that area includes  the solution near a plane   wall~\cite{blake1971note,BlakeChwang,spagnolie_lauga_2012}, near a sphere~\cite{Oseen,SphereStokeslet,Alex} or between two infinite plates~\cite{Liron1976}. 
	{In essentially all these cases,    the presence of a surface leads to a flow induced at the location of the point force, which means that the forced rigid body  experiences an additional surface-induced advection.}
	In this paper, we consider a point force located outside of a rigid, slender filament {and propose a theoretical approach to compute the filament-induced advection of the point force}. 
	
	The primary motivation to consider a point force near a filament stems from a classical biological situation. 	As part of the normal function of eukaryotic cells, various cargoes require to be transported in the cell's cytoskeleton,  a fibrillar mesh composed of different elongated slender polymer filaments, from soft actin to more rigid microtubules~\cite{bershadsky2012cytoskeleton}. This form of intracellular transport is powered by motor proteins (myosins, kinesins and dyneins) that ratchet their way along the filaments and drag their cargo along with them~\cite{transportVale,transportWang}. Since this transport takes place inside a cell, and thus immersed in the viscous cytoplasm~\cite{cytoplasm}, it is  important  to quantify the hydrodynamic interactions between a forced rigid body (cargo) and a slender filament (polymer).

	In this paper, we outline a physically-intuitive approximation for
	the flow in the limit of an intermediate   separation between the slender filament and the point force. We assume that the distance between the filament and the point force (denoted by $d$) is large compared to the diameter of the cross-section of the filament (denoted by $2a$), but   small compared to its length ($2L$), so we are in the limit $a \ll d\ll L$. 	The full solution for a point force  outside an infinite cylinder is already known~\cite{Full_calculation} but it can be used as an approximation in practice only for extremely slender filaments. Here we thus seek a different approach that will provide us with a better approximation for filaments with finite  aspect ratios.

	The literature on the response of slender filaments to external flows is substantial. The majority of the past theoretical results relies on the well-known framework of slender-body theory (SBT)~\cite{Batchelor1970419,Cox1970791,Clarke1972781,keller_rubinow_1976,Lighthill1976161,Johnson1980411,Gotz2000,Koens2016,Koens2017}. Built on the idea of approximating the flow around a slender filament by a distribution of hydrodynamic singularities along its centreline (primarily point forces and source dipoles), this framework can be thought of as an asymptotic approximation of the boundary integral formulation of the Stokes flow~\cite{pozrikidis_1992,koens_lauga_2018}. 
	SBT has been of great use in modelling   the suspensions of passive slender fibres~\cite{Tornberg20048,Tornberg2006172} as well as the swimming of elongated microorganisms~\cite{Barta1988992,Koens2014,Myerscough1989201,Smith2009289}. {The combined dynamics of passive and active filaments is relevant to}  the motion of microswimmers through a matrix of rigid obstacles~\cite{Shelley2012,Shelley2014,Chamolly_2017,theresa2019,Guasto2019}. 
	
	In this paper, we propose a physically-intuitive approach {to compute the impact of the filament on the velocity of the point force itself}  that lays the groundwork to  investigate  the interactions between higher-order hydrodynamic singularities and slender filaments of complex shapes. This will allow for future studies on the motion of microswimmers~\cite{Lighthill1975,Wu1975,Bray2001,Berke2008,Lauga2009,Drescher201110940,Lauga2016105,Saintillan2018563} through filamentous media, such as cervical mucus~\cite{mucus}. After introducing the setup of the problem (\S~\ref{sec::setup}), we show that the flow can be captured by the use of adjusted local resistive-force theory coefficients (\S~\ref{sec::drag}), which  we then use  to predict the advection of the point force in the presence of the filament (\S~\ref{sec::advection}). We next validate our theoretical results using boundary element method computations (\S\ref{sec::numerics}) and investigate how to extend the results for curved filaments (\S~\ref{sec::curved}). Our numerical code is validated in Appendix~\ref{sec:validate}.

	\section{Setup and General framework \label{sec::setup}}
	
	\begin{figure}[t]
		\includegraphics[width=0.65\linewidth]{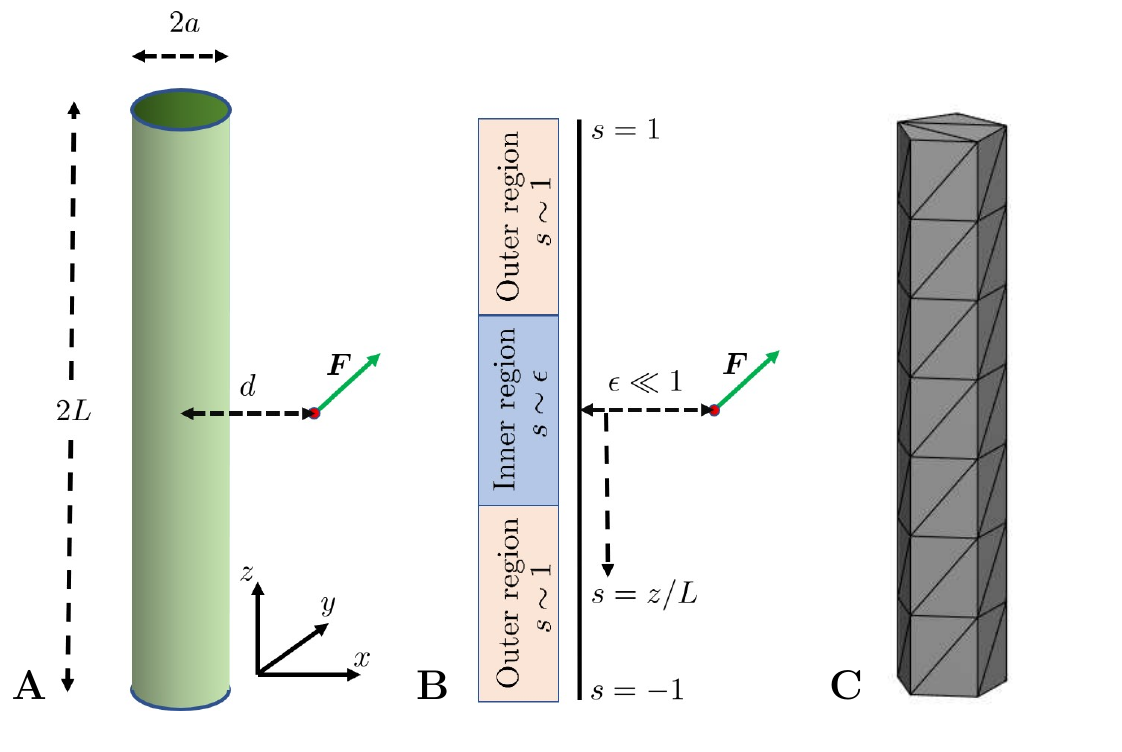}
		\caption{A: Schematic illustration of our setup.  A point force is located at a distance $d$ from a centreline of the cylinder of diameter $2a$ and length $2L$ and we derive the flow in the limit $a \ll d\ll L$. B: Schematic illustration of the setup with lengths scaled by $L$ and with the regions relevant for the asymptotic analysis labelled. C: Mesh used for the boundary element computations.}\label{fig::setup_fig}
	\end{figure}
	
	The setup consists of a fixed, rigid, circular cylinder of radius $a$ and length $2L\gg 2a$ as illustrated in Fig.~\ref{fig::setup_fig}A. A point force of strength $\bm{F}$ is set at a distance $d\gg a$ from the axis of the cylinder and contained in the plane of the middle cross-section of the cylinder. For the purpose of calculations, we introduce a set of Cartesian coordinates  with the origin at the location of the singularity and the axis of the cylinder described by the  line segment $y=0, x=-d, -L\leq z \leq L$ (see Fig.~\ref{fig::setup_fig}A). 
	
	The  system is immersed in an incompressible, Newtonian fluid of dynamic viscosity $\mu$ that is otherwise at rest. Motivated by  applications from biology and complex fluids, we focus on the low-Reynolds number limit and  assume that the  flow is governed by the incompressible Stokes equations. Thus, we solve for a velocity field $\bm{u}$ and a pressure field $p$ such that 
	\begin{eqnarray}
		-\mu \bm{\nabla}^2 \bm{u} + \bm{\nabla} p = \bm{F} \delta(\bm{x}), \, \, \bm{\nabla}\cdot\bm{u} = 0, \\
		\bm{u}|_{\Omega_c} = \bm{0} \text{ and } |\bm{u}(\bm{x})| \to 0 \text{ as } |\bm{x}|\to \infty,  
	\end{eqnarray}
	where $\Omega_c$ is the surface of the cylinder.
	
	In the absence of the cylinder, the solution for the flow due to a point force singularity, 
	\begin{equation}
		\bm{u}^{(0)} = \frac{1}{8\pi\mu} \left(\frac{\mathbb{I}}{r}+\frac{\bm{x}\bm{x}}{r^3}\right)\cdot \bm{F}\triangleq \bm{G}(\bm{x})\cdot \bm{F}, \label{eq::stokeslet}
	\end{equation}
	where $\mathbb{I}$ is the identity tensor, satisfies all equations except for the no-slip condition on the surface of the cylinder. Rather than determining the total flow, it suffices then to derive the reaction flow $\bm{u}^{(1)} = \bm{u}-\bm{u}^{(0)}$ due to the presence of the cylinder that solves the incompressible Stokes equations with no {long-range volume forces acting on the fluid} and with $\bm{u}^{(1)} = -\bm{u}^{(0)}$ on the surface of the cylinder. 
	
	Since the cylinder has a large aspect ratio, $L\gg a$, we can exploit the framework of   slender-body theory (SBT)~\cite{Batchelor1970419,Cox1970791,Clarke1972781,keller_rubinow_1976,Lighthill1976161,Johnson1980411,Gotz2000,Koens2016,Koens2017} to find the leading-order value (in $a/d$) of the reaction flow as a flow due to a distribution of point forces $\bm{f}$ set along the axis of the cylinder, as  
	\begin{eqnarray}
		&&\bm{u}^{(1)}(\bm{y}) = \int_{-L}^{L} \bm{G}[\bm{y}-\bm{x}(z)] \cdot \bm{f}(z) \,dz, \label{SBT_flow}
	\end{eqnarray}
	with $\bm{x}(z) = (-d,0,z)$ parametrising the axis of the cylinder and $\bm{y}$ being any point in the fluid domain such that its distance to the filament is much larger than $a$.
	We exploit the fact that the cylinder is much longer than the distance to the singularity $\epsilon \triangleq d/L\ll 1$ and perform an asymptotic calculation in   $\epsilon$ to adjust the classical resistive-force theory~\cite{Hancock_stokeslet,GRAY775,Lighthill1976161}, which is the  local form of the SBT, to this setting.

	\section{Adjusted resistive-force theory \label{sec::drag}}
	Starting from the version of slender-body theory derived by Keller and Rubinow~\cite{keller_rubinow_1976}, we can write the integral equation   for the  force distribution {$\bm{f}(z)$ along the axis of the cylinder} as
	\begin{eqnarray}
		-8\pi\mu\bm{u}^{(0)}(z) = [(L_{S}+1)(\mathbb{I}-\hat{\bm{z}}\hat{\bm{z}})+2(L_{S}-1)\hat{\bm{z}}\hat{\bm{z}}]\cdot\bm{f}(z)  +(\mathbb{I}+\hat{\bm{z}}\hat{\bm{z}})\int_{-L}^{L} \frac{\bm{f}(z')-\bm{f}(z)}{|z'-z|} \,dz'\, , \label{SBT}
	\end{eqnarray} 
	with $L_{S} = \ln 4(L^2-z^2)/a^2$ and where $\bm{u}^{(0)}(z)$ is the flow velocity due to a point force set at the origin evaluated at the axis of the cylinder i.e.~at the point $-d\hat{\bm{x}}+z\hat{\bm{z}}$. 
	{
		Note that this force   distribution $\bm{f}(z)$ is the same as the one appearing in  Eq.~\eqref{SBT_flow}}.	Since the cylinder is straight, the directions are decoupled and Eq.~\eqref{SBT} becomes
	\begin{subeqnarray}
		-8\pi \mu u^{(0)}_k(z) &=& (L_{S}+1) f_k(z)+\int_{-L}^{L} \frac{f_k(z')-f_k(z)}{|z'-z|} \,dz', \\
		-4\pi \mu u_z^{(0)}(z) &=& (L_{S}-1) f_z(z)+\int_{-L}^{L} \frac{f_z(z')-f_z(z)}{|z'-z|} \,dz',
	\end{subeqnarray}
	with $k=x,y$. We focus on the equation in the $x$ direction and drop the $x$ subscript as the other directions follow similarly. After introducing $\alpha = 1/(L_{S}+1)$ and $f_0(z) = -8\pi \mu \alpha u^{(0)}_x(z)$, the integral equation for $f(z)$ becomes
	\begin{equation}\label{integraleq}
		f_0(s) = f(s)+\alpha \int_{-1}^{1} \frac{f(s')-f(s)}{|s'-s|} \,ds',
	\end{equation}
	where we introduced {the} non-dimensional arclength along the axis $s =z/L$ to facilitate the asymptotic analysis.  
	
	The common way to solve this Fredholm integral equation of the second kind is to use the iteration scheme under the assumption $\alpha \ll 1$ with $f = f_0$ as initial guess~\cite{keller_rubinow_1976}. This   guess is referred to as the resistive-force theory (RFT), an approximation that excludes nonlocal hydrodynamic interactions between  different cross-sections of the cylinder. In our setting, we have another small quantity $\epsilon = d/L$, the ratio between the cylinder-singular distance and the cylinder (half) length. On the left-hand side of Eq.~\eqref{integraleq}, $f_0$ has a functional form proportional to $(\epsilon^2+s^2)^{-1/2}$ as the stokeslet singularity decays as the inverse distance from the singularity. Assuming that $f(s)$ has a similar behaviour (which we will check a posteriori), we can compare the contributions from the ``outer'' $s'\sim \mathcal{O}(1)$ and ``inner'' $s'\sim  \mathcal{O}(\epsilon)$ regions to the integral on the right-hand side of Eq.~\eqref{integraleq}, when $s\sim \mathcal{O}(\epsilon)$ i.e.~when solving for the force distribution in the ``inner'' region (see Fig.~\ref{fig::setup_fig}B). We focus on this ``inner'' region  because our goal is to evaluate the flow at the position of the singularity i.e.~integrate the flows of the point forces $\bm{f}\,ds$ propagated by the stokeslet singularity. This results in the integration of a function of the form $(\epsilon^2+s^2)^{-1}$, which clearly has a dominant contribution coming from the $s\sim \mathcal{O}(\epsilon)$ region.
	
	Next, the numerator  $f(s')-f(s)$ will always be on the order of $\epsilon^{-1}$, since when $s'\sim  \mathcal{O}(1)$ we have $f_0(s')\sim \mathcal{O}(1)$. This type of integrals is well-known~\cite{hinch_PM} and often appears in slender-body calculations. The main contribution comes from the intermediate region with $\epsilon\ll s' \ll 1$ where $f(s')-f(s)\approx -f(s)$. Hence, we need to integrate $-f(s)/|s'-s|$, which is again a well known result~\cite{hinch_PM} and we finally obtain
	\begin{equation}
		f_0(s) \sim f(s)-2\alpha f(s) \ln 1/\epsilon+\mathcal{O}\left( \frac{\epsilon^{-1}}{\ln^{2} L/a} \right).
	\end{equation} 
	We can therefore state  that
	\begin{eqnarray}
		f_x(s)\sim -\frac{8 \pi \alpha u_x(s)}{1-2\alpha \ln 1/\epsilon} = - \frac{8\pi u_x(s)}{1+L_{S}-2\ln L/d} \approx - \frac{8\pi u_x(s)}{1+2\ln 2d/a},
	\end{eqnarray}
	where the last approximation is made in $L_S\approx 2 \ln (2L/a)$, valid as $s\ll 1$.
	Thus, the RFT for the slender filament  is  ``adjusted'' through the presence of  new perpendicular and parallel drag coefficients
	\begin{subeqnarray}\label{eq:newdragcoef}
		\tilde{\xi}_{\perp} &=& \frac{4\pi\mu}{\ln ({2d}/{a})+0.5}, \label{perp_drag}\\
		\tilde{\xi}_{\parallel} &=& \frac{2\pi\mu}{\ln ({2d}/{a})-0.5}. \label{par_drag}
	\end{subeqnarray}
	These coefficients have the same form as in the standard RFT~\cite{Cox1970791} but with the effective length of the cylinder replaced by the distance $d$. 
	
	\section{Advection of the point force \label{sec::advection}}
	
	To determine how the singularity interacts with the cylinder, we next write the reaction flow of the cylinder   of a line of point forces distributed along its axis as in Eq.~\eqref{SBT_flow} and whose magnitude $\bm{f}$ is set by the effective drag coefficients in 
	Eqs.~\eqref{perp_drag} above. The effective drag coefficients are only valid in the ``inner'' region $z\sim d$ (i.e.~$s\sim \epsilon$ in Fig.~\ref{fig::setup_fig}B). However, because of the fast spatial decay of both the stokeslet and the force distribution on the cylinder, the main contribution to the flow originates from this $z\sim d$ region and it is therefore appropriate to use the adjusted drag coefficients in determining the flow at the location of the point force. We thus write
	\begin{eqnarray}
		&&\bm{u}_S^{(1)}\sim -\frac{1}{8\pi\mu} \int_{-L}^{L} \left(\frac{\mathbb{I}}{r}+\frac{\bm{x}\bm{x}}{r^3}\right)\cdot [\tilde{\xi}_{\perp}(\mathbb{I}-\hat{\bm{z}}\hat{\bm{z}})+ \tilde{\xi}_{\parallel}\hat{\bm{z}}\hat{\bm{z}}]\cdot \bm{u}^{(0)}(\bm{x}) \, dz,
	\end{eqnarray}
	where we use  the subscript $S$  to indicate that it is the velocity of the reaction flow at the location of the singularity. Since   $\bm{x}(z) = (-d,0,z), r = |\bm{x}|$ and $\bm{u}^{(0)}(\bm{x}) = \bm{G}(\bm{x})\cdot \bm{F}$ where $8\pi\mu\,\bm{G}(\bm{x}) = \mathbb{I}/r+\bm{x}\bm{x}/r^3$ is the Oseen tensor, we may further write
	\begin{eqnarray}
		&&\bm{u}^{(1)}_S \sim -\frac{\bm{F}}{8\pi\mu d} \cdot \int_{-L/d}^{L/d} \left(\frac{\mathbb{I}}{r}+\frac{\bm{x}\bm{x}}{r^3}\right)\cdot \left[\frac{\tilde{\xi}_{\perp}}{8\pi\mu}(\mathbb{I}-\hat{\bm{z}}\hat{\bm{z}})+ \frac{\tilde{\xi}_{\parallel}}{8\pi\mu}\hat{\bm{z}}\hat{\bm{z}}\right]\cdot \left(\frac{\mathbb{I}}{r}+\frac{\bm{x}\bm{x}}{r^3}\right) \, d\zeta  , \label{eq::advection_integral_form}
	\end{eqnarray}
	with $\zeta = z/d$ the rescaled integration variable, appropriate for the $z\sim d$ part of the filament with the dominant contribution to the reaction flow. Note that in Eq.~\eqref{eq::advection_integral_form} and in what follows, we non-dimensionalise all   spatial variables by $d$ but kept the same symbols for notation simplicity.  
	
	As expected from the linearity of the Stokes equations and  dimensional analysis, we can write $\bm{u}^{(1)}_S = -(8\pi \mu d)^{-1}\,\bm{F}\cdot\bm{M}$ for some dimensionless second-rank tensor $\bm{M}(a/d,L/d)$, which, by symmetry,   is diagonal in the $(x,y,z)$ frame. 
	It is represented by the dimensionless integral in Eq.~\eqref{eq::advection_integral_form}, and thus the diagonal components of $\bm{M}$ are
	\begin{subeqnarray}
		M_{xx} &=& \frac{\tilde{\xi}_\perp}{8\pi\mu} \int_{-\infty}^{\infty}d\zeta\,\frac{4+\zeta^2}{r^4} + \frac{\tilde{\xi}_\parallel-\tilde{\xi}_\perp}{8\pi\mu} \int_{-\infty}^{\infty}d\zeta\,\frac{\zeta^2}{r^6} \nonumber \\
		&=& \frac{\pi(19+\Gamma)}{8[1+2\ln (2d/a)]} { + \mathcal{O}\left(\frac{d/L}{\ln L/a},\frac{1}{\ln^2 L/a}\right)}, \label{Mxx}\\ 
		M_{yy} &=& \frac{\tilde{\xi}_\perp}{8\pi\mu} \int_{-\infty}^{\infty} \frac{d\zeta}{r^2} = \frac{\pi}{1+2\ln (2d/a)} { + \mathcal{O}\left(\frac{d/L}{\ln L/a},\frac{1}{\ln^2 L/a}\right)}, \\
		M_{zz} &=& \frac{\tilde{\xi}_\perp}{8\pi\mu} \int_{-\infty}^{\infty}d\zeta\,\frac{1+4\zeta^2}{r^4} + \frac{\tilde{\xi}_\parallel-\tilde{\xi}_\perp}{8\pi\mu} \int_{-\infty}^{\infty}d\zeta\, \frac{(1+2\zeta^2)^2}{r^6}\nonumber \\ 
		&=& \frac{\pi(1+19\Gamma)}{8[1+2\ln (2d/a)]} { + \mathcal{O}\left(\frac{d/L}{\ln L/a},\frac{1}{\ln^2 L/a}\right)},
	\end{subeqnarray}
	with the ratio of drag coefficients $\Gamma$ given by 
	\begin{equation}
		\Gamma = \frac{\tilde{\xi}_\parallel}{\tilde{\xi}_\perp} = \frac{1}{2}\frac{\ln (2d/a)+0.5}{\ln (2d/a)-0.5}.
	\end{equation}
	In the asymptotically-slender limit $\ln d/a \gg 1$, we can approximate $\Gamma \approx 1/2$ and retrieve the   results from Ref.~\cite{Full_calculation}, which are obtained using an asymptotic expansion of the exact solution for the flow induced by a point force near a rigid infinite cylinder ($L \to \infty$) in the $d/a\gg1$ limit.

	\begin{figure*}[t]
		\includegraphics[width=\linewidth]{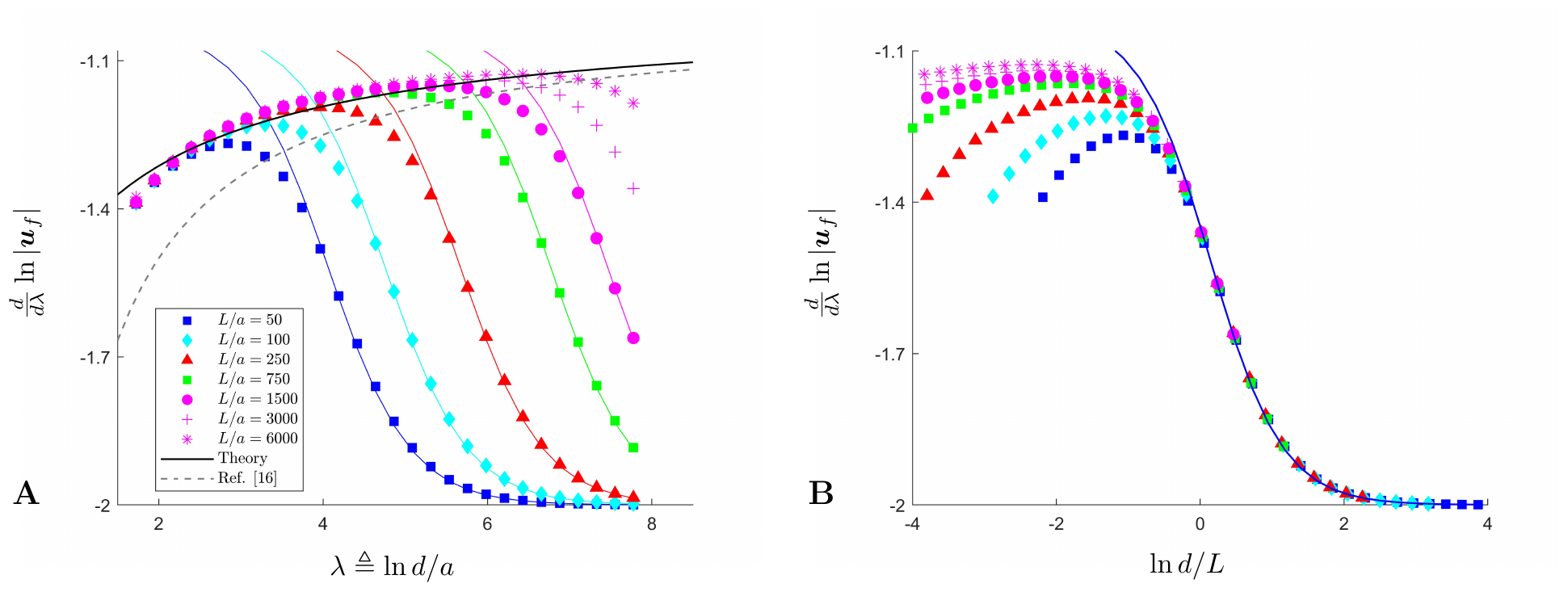}
		\caption{A: Decay gradient of the reaction flow for various cylinder aspect ratios $L/a$ {in the case of a point force pointing in the positive $x$ direction}. The numerical results of boundary element method (BEM) computations are shown by symbols while the predictions from the standard RFT with drag coefficients independent of $d${, shown in Eq.~\eqref{eq::rft}}
			follow the appropriately coloured solid lines. We plot our theoretical prediction in Eq.~\eqref{decay_theory}   as a solid black line while the infinite cylinder prediction from Ref.~\cite{Full_calculation} is shown as a grey dashed line. {Computations shown with filled symbols were performed with the same uniform size of surface elements while the ones represented by (+) and ($\ast$) have  elements that are 2 and 4 times larger, respectively, in the direction of the axis of the cylinder.} B: The same data plotted against $\ln d/L$. The RFT prediction lines from panel A all collapse onto the single  solid blue line in  in panel B.}\label{fig::main_fig}
	\end{figure*}
	
	\section{Comparison with numerical computations \label{sec::numerics}}

	To  test our theoretical results, we use simulations based on the boundary element method (BEM)~\cite{Jaswon196323,Symm1963,Brebbia1978,Pozrikidis20021}. We mesh the cylinder into uniform triangles with 5 cross-sectional vertices (see Fig.~\ref{fig::setup_fig}C).  A point force of unit strength is then placed at a distance $d$ and we solve for the single layer potential on the surface of the cylinder corresponding to the reaction flow $\bm{u}^{(1)}$. This allows us to evaluate the reaction flow at the location of the point force, $\bm{u}^{(1)}_S$, and to examine its spatial decay with a change in the value of $d$. Our numerical code is validated in Appendix~\ref{sec:validate} against two exact solutions (point force near a rigid surface and outside a sphere);
	{we also verify there that changing the number of cross-sectional vertices does not   significantly impact our  results.}

	In Fig.~\ref{fig::main_fig}A we show the results of numerical computations in the form of $d \ln |\bm{u}_S^{(1)}| / d \lambda$ versus $\lambda = \ln (d/a)${, for a point force pointing in the positive $x$ direction}. This quantity represents the slope of the log-log {dependence} of the advection   of the point force as a function of the separation $d$ between the point force and the cylinder. We choose to plot this quantity for two reasons. First, a power-law decay of the form $\bm{u}_S^{(1)}\propto d^{-k}$ has a constant log-log slope $d \ln |\bm{u}_S^{(1)}| / d \lambda \equiv -k$. In the limit of  large separations $d\gg L$ we know that the point force and the cylinder will interact as stokeslets with $1/d$ decay, so the advection rate decays as $1/d^2$ and we expect a plateau to $d \ln |\bm{u}_S^{(1)}| / d \lambda \to -2$. Secondly, a common way of visualising the information about spatial decay is to   show the log-log plot of $|\bm{u}_S^{(1)}|$ and $d$. However, the variation in the slope would not be   prominent in such a plot since we are interested in the initial increase in the slope that varies from -1.4 to -1.2 (see Fig.~\ref{fig::main_fig}A).

	In the large separation limit $d\gg L$, the reaction flow is well approximated by a point force and thus the $d^{-2}$ decay of $|\bm{u}_S^{(1)}|$ is expected. We do recover this prediction on the far-right of Fig.~\ref{fig::main_fig}A with $d \ln |\bm{u}_S^{(1)}|/d \lambda$ converging to $-2$. 
	The   approach to   this plateau over the range of separations $d\gtrsim L$ is well captured by the standard RFT with drag coefficients  independent of the separation $d$, given by  \cite{Cox1970791}
	{
		\begin{equation}
			{\xi}_{\perp} = \frac{4\pi\mu}{\ln ({2L}/{a})+0.5}, \hspace{2cm} {\xi}_{\parallel} = \frac{2\pi\mu}{\ln ({2L}/{a})-0.5}. \label{eq::rft}
		\end{equation}
		The theoretical predictions based on the standard RFT are obtained with a calculation based on Eq.~\eqref{eq::advection_integral_form} where $\xi_{\parallel}$ and $\xi_{\perp}$ from Eq.~\eqref{eq::rft} are used instead of $\tilde{\xi}_{\parallel}$ and $\tilde{\xi}_{\perp}$ from Eq.~\eqref{eq:newdragcoef}. The results are  shown in Fig.~\ref{fig::main_fig}A as solid lines of colour (we use the same colours as the computations).} Furthermore, we are able to collapse these large-$d$ theoretical predictions   by shifting the horizontal axis by $\ln L/a$, as shown in  Fig.~\ref{fig::main_fig}B. Indeed, for large values of $d/L$, the hydrodynamics become independent of the radius of the cylinder $a$ and so the decay is governed by $d/L$ or $\lambda - \ln L/a = \ln d/L$ in the plot.

	In the case of  intermediate values of $d$ with $a\ll d\ll L$, our theoretical result in Eq.~\eqref{Mxx} 
	with the point force in the $x$ direction leads to the prediction
	\begin{equation}
		\frac{d \ln |\bm{u}_S^{(1)}|}{d \lambda} = -1 - \frac{1}{\lambda +\ln 2+0.5},
		\label{decay_theory}
	\end{equation} 
	where we have assumed that the ratio of drag coefficients takes its slender value, i.e.~$\Gamma\approx 1/2$. The solid black line in  Eq.~\eqref{decay_theory} shows this new prediction,  which is  in good agreement with the computational results.

	In the same plot we also show as a dashed line the theoretical prediction based on the asymptotic analysis of the   solution with  an infinite cylinder~\cite{Full_calculation}. Our theoretical prediction   lead to a much better agreement with the BEM computations in the case a finite-size filament than the result for infinite cylinder. The  work presented in this paper allows therefore to capture the hydrodynamic interactions between a point force and a slender  filament of finite size.  
	
	Our computational results may be extended to the case where the point force is not placed symmetrically to be equidistant from both ends of the cylinder (i.e.~not in the $xy$ plane, see Fig.~\ref{fig::setup_fig}A).  Since the dominant contribution to the flow induced at the location of the singularity arises from the part of the filament that is closest to the point force, if the displacement of the point force in the $z$ direction is comparable with the separation $d$, the flow should not change at  leading order.  BEM computations carried out with  point force displaced along $z$ by  up to $10a$ show that, for the same range of $d/a$ as in Fig.~\ref{fig::main_fig}A,   the values of  $d \ln |\bm{u_S^{(1)}}| / d \lambda$ stay within 1\% of the original case  (not shown).

	\section{Curved filaments \label{sec::curved}}
\begin{figure*}[t]
		\includegraphics[width=\linewidth]{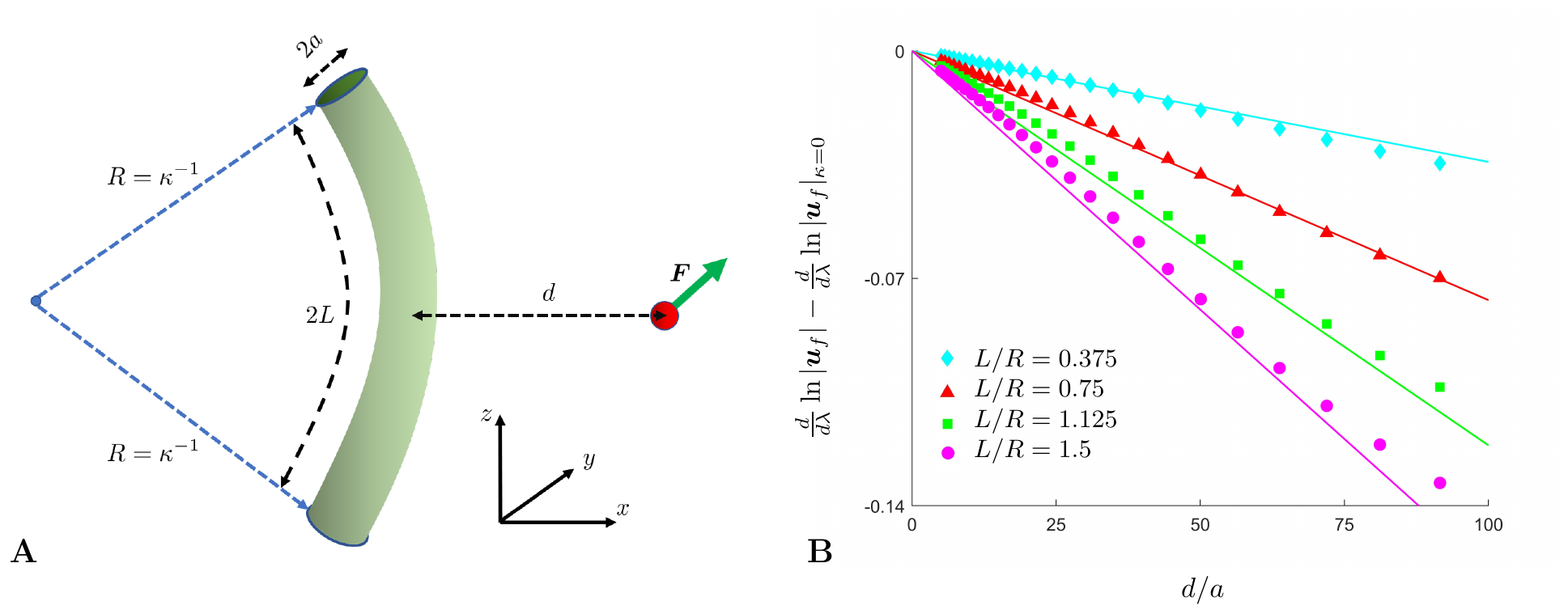}
		\caption{A: Setup for the calculation in the case of a curved filament. B: Correction to the decay rate due to the curvature as a function of the separation $d/a$, for various relative curvatures $\alpha = L/R$. {The point force is pointing in the positive $x$ direction.} Symbols represent results of the BEM computations and solid lines follow the theoretical prediction from Eq.~\eqref{theory_curved}.}\label{fig::curvature}
	\end{figure*}

	For the asymptotic calculations proposed above to hold, the filament has to have a large aspect ratio. In an experiment, such a slender filament would be flexible and  prone to deformations, so it is important to investigate how a more complex slender shape for the filament centreline interacts with a nearby point force. {Here we present preliminary results in the case of the simplest   choice for a  curved shape, namely}  a uniformly curved filament with its centreline forming a circular arc of radius $R = \kappa^{-1}$ (see Fig.~\ref{fig::curvature}A).

	We assume that the filament is curved in such a way that it stays in the $xz$ plane and that its  ends are further away from the point force than its midpoint; this ensures that the midpoint is always the part of the filament closest to the force so that we can rely on our calculations above for   straight filaments. The centreline of the filament is parametrised by the arclength $s$ via $\bm{x} = [-d-R(1-\cos (s/R)),0,R \sin(s/R)]$.  Apart {from}  the curvature, all   other parameters of the problem remain the same as in the previous section.

	We may repeat the  calculation from above and determine the  rate  at which the point force is advected by the presence of the filament, $\bm{u}_S^{(1)}$, i.e.~the reaction flow at the location of the force. These calculations are based on using the RFT for approximating the reaction flow, but with adjusted drag coefficients, as in Eqs.~\eqref{perp_drag},   with
	\begin{eqnarray}
		\bm{f} = -[\tilde{\xi}_{\perp}(\mathbb{I}-\hat{\bm{t}}\hat{\bm{t}})+ \tilde{\xi}_{\parallel}\hat{\bm{t}}\hat{\bm{t}}]\cdot \bm{u}^{(0)}(\bm{x}),
	\end{eqnarray}
	where $\bm{u}^{(0)} = (8\pi\mu)^{-1} (\mathbb{I}/r+\bm{x}\bm{x}/r^3)\cdot \bm{F}$ is  the bulk stokelset flow and $\bm{\hat{t}} = [-\sin(s/R),0,\cos (s/R)]$ the unit tangent along the centreline of the filament. As above, we aim to compute the diagonal entries of the non-dimensional matrix $\bm{M}$, defined through $\bm{u}_S^{(1)} = -(8\pi\mu d)^{-1} \bm{F}\cdot\bm{M}$, which,  within the RFT framework, can be approximated by the dimensionless integral
	\begin{equation}
		\bm{M} = \int_{-L/d}^{L/d} \left(\frac{\mathbb{I}}{r}+\frac{\bm{x}\bm{x}}{r^3}\right)\cdot \left[\frac{\tilde{\xi}_{\perp}}{8\pi\mu}(\mathbb{I}-\hat{\bm{t}}\hat{\bm{t}})+ \frac{\tilde{\xi}_{\parallel}}{8\pi\mu}\hat{\bm{t}}\hat{\bm{t}}\right]\cdot \left(\frac{\mathbb{I}}{r}+\frac{\bm{x}\bm{x}}{r^3}\right) \, d\zeta , \label{eq::curved_M_integral}
	\end{equation}
	with $\zeta = s/d$ and with all the other spatial variables non-dimensionalised by $d$. This integral can be determined analytically in a closed form but the full expression is cumbersome and thus we do not include it.    Instead, we note that by introducing another length scale to the problem, $R$, we can form two additional dimensionless groups that describe the setup, namely $\alpha = L/R$ and $\beta = d/R$. We assume that the radius of curvature is comparable with the length of the cylinder, $\alpha \sim \mathcal{O}(1)$ and thus $\beta \ll 1$. Under these assumptions, we expect the corrections due to the curvature to be small compared to the case of a straight filament since the main contribution arises from the $s\sim d \ll R$ part of the filament, which is only very slightly curved. To determine these corrections, we expand   Eq.~\eqref{eq::curved_M_integral} in powers of $\beta$ and obtain explicitly  the leading-order term in $\beta$ to be
	\begin{subeqnarray}
		M_{xx} &=& M_{xx}|_{\kappa=0}-\frac{\tilde{\xi}_\perp}{8\pi\mu}\frac{\beta}{32} \left(15 \pi -12 \alpha +36 \sin \alpha +32 \cot \frac{\alpha
		}{2} \right)+\mathcal{O}(\beta^2),  \label{Mxx_curved}\\ 
		M_{yy} &=& M_{yy}|_{\kappa=0}-\frac{\tilde{\xi}_\perp}{16\pi\mu}\beta\left(\pi+2\cot \frac{\alpha}{2} \right) +\mathcal{O}(\beta^2), \\
		M_{zz} &=& M_{zz}|_{\kappa=0} - \frac{\tilde{\xi}_\perp}{8\pi\mu}\frac{\beta}{32} \left(69 \pi-12 \alpha -36 \sin \alpha + 64 \cot
		\frac{\alpha }{2}\right)+\mathcal{O}(\beta^2),\label{eq:last}
	\end{subeqnarray}
	where again we assumed $\Gamma = \tilde{\xi}_{\parallel}/\tilde{\xi}_{\perp} = 1/2$.

	We plot in  Fig.~\ref{fig::curvature}B  the results of the BEM computations for a filament of fixed aspect ratio and with various radii of curvature, performed similarly to what was described for straight filaments. 	The point force is in the $x$ direction and we plot the difference in the previously introduced quantity,  $d \ln |\bm{u_S^{(1)}}| / d \lambda$, between the values for the curved and straight filament. The result in Eq.~\eqref{Mxx_curved} provides us with a theoretical prediction as 
	\begin{equation}
		\frac{d}{d \lambda}\ln |\bm{u}_f|-\frac{d}{d \lambda}\ln |\bm{u}_f|_{\kappa=0} \approx -\frac{a}{78 R \pi} \left(15 \pi -12 \alpha +36 \sin \alpha +32 \cot \frac{\alpha
		}{2} \right) \times  \frac{d}{a}, \label{theory_curved}
	\end{equation}
	which is a linear function of $d/a$ with a slope that depends on $\alpha = L/R$ and $R/a$. This theoretical prediction is plotted for a filament of aspect ratio $L/a = 250$ and different values of $\alpha$ and these are shown as appropriately coloured solid lines in Fig.~\ref{fig::curvature}B. The theoretical prediction from Eq.~\eqref{theory_curved} agrees well with the BEM computations over a wide range of dimensionless curvatures.

	Note that the corrections to the tensor $\bm{M}$   in Eqs.~\eqref{Mxx_curved} as an expansion in $\beta\ll 1$ are valid only within the adjusted RFT approximation i.e.~the RFT with drag coefficients given by Eqs.~\eqref{perp_drag}. The adjusted RFT is an exact asymptotic approximation for the force distribution only in the ``inner'' region $s\sim d$; however, in calculating the $\mathcal{O}(\beta)$ terms in Eqs.~\eqref{Mxx_curved}, we kept the finite limits $\pm L/d$ of the integral in Eq.~\eqref{eq::curved_M_integral}. We thus included the subdominant contribution of the ``outer'' region $s\sim L$ but with the drag coefficients from the ``inner'' region;  keeping the integral limits finite gives a significantly better agreement in  	Fig.~\ref{fig::curvature} with the results of BEM computations. 
	
	{Although the above are   preliminary results, they show convincingly that the method developed in this paper  also works in the case of non-straight filaments; further work will be able to quantify in which limit this approach breaks down.}

	\section{Discussion \label{sec::discussion}}
	
	In this article, we investigate the reaction flow outside a rigid, slender filament due to a point force located outside of it at an intermediate distance (i.e.~at a distance intermediate between the filament radius and length). We exploit the slenderness of the filament to derive a physically-intuitive approximation of the flow, through the framework of the slender-body theory. An asymptotic analysis of the problem reveals a leading-order approximation for the force distribution along the axis of the filament in the form of an adjusted resistive-force theory (RFT) with drag coefficients that are not intrinsic to the filament but   depend on its distance to the point force. {We next use these new drag coefficients to compute the filament-induced advection of the point force}. 
	We then test this theoretical prediction against boundary element computations and obtain excellent agreement. Using this adjusted RFT, we next form a theoretical prediction on how the flow would change if the filament is curved   and again find good agreement with computations.   	While a full solution was already known for the flow due to a point force outside an infinite cylinder~\cite{Full_calculation},   	the theory presented in this letter provides significantly better predictions for large but finite aspect ratio filaments. With our new  approximation for the flow, it will be   easier to derive higher moments of the reaction flow at the location of the force, such as torque, stress etc.  	Our results    could also be generalised for higher-order hydrodynamic singularities and thus allow to investigate the interactions between     microswimmers and  biological filaments.

	\section{Acknowledgements}
	This project has received funding from the European Research Council under the European Union's Horizon 2020 Research and Innovation Programme (Grant No. 682754 to E.L.) and from   Trinity College, Cambridge (IGS scholarship to I.T.).

	\appendix
	{\section{Code validation}\label{sec:validate}
		To confirm the validity of our BEM code, we show here its convergence to  exact solutions in two classical cases: a point force near a rigid plane wall~\cite{blake1971note} and near a rigid sphere~\cite{KimKarrila,SphereStokeslet,Oseen}. {Additionally, we   demonstrate the robustness of the code  to different cross-sectional geometries used on the same setup as in the main body of the paper.}
		
		The total flow  due to a point force $\bm{F}$ located at $\bm{y}$ in these two {test} geometries can be written as $\bm{G}(\bm{x}-\bm{y})\cdot\bm{F}+\bm{u}_E(\bm{x})$, where $\bm{x}$ is any point in the fluid domain and $\bm{G}$ is the Green's function defined in Eq.~\eqref{eq::stokeslet}. In both geometries, the reaction (image) flow has a closed analytic form $\bm{u}_E(\bm{x})$ that {in most cases} can be expressed as a finite sum of hydrodynamic singularities set at an image point $\bm{y}^*$ (below the surface~\cite{blake1971note} or inside the sphere~\cite{KimKarrila}). The exact functional dependence of $\bm{y}^*$ on $\bm{y}$ depends on the particular geometry of the solid boundary and we define $\bm{r} = \bm{x}-\bm{y}^*$ and $r = |\bm{r}|$ for any point $\bm{x}$ in the fluid domain.

		In these standard geometries, we can   compare the results  of our numerical computations $\bm{u}_N$ against the exact solution $\bm{u}_E$, thereby testing the validity of our BEM implementation. We choose an appropriate finite set of points $S_f$ in the fluid domain and formulate a distance between the numerical solution $\bm{u}_N$ and the exact solution $\bm{u}_E$ as a root mean squared (RMS) error   as a square root of
		$
		|S_f|^{-1}\sum_{\bm{x}\in S_f}|\bm{u}_N(\bm{x})-\bm{u}_E(\bm{x})|^2.
		$
		Henceforth, this RMS error  will be used to demonstrate the convergence of the numerical results towards the exact solution as the surface meshing gets refined.
		
		\begin{figure*}[t]
			\includegraphics[width=\linewidth]{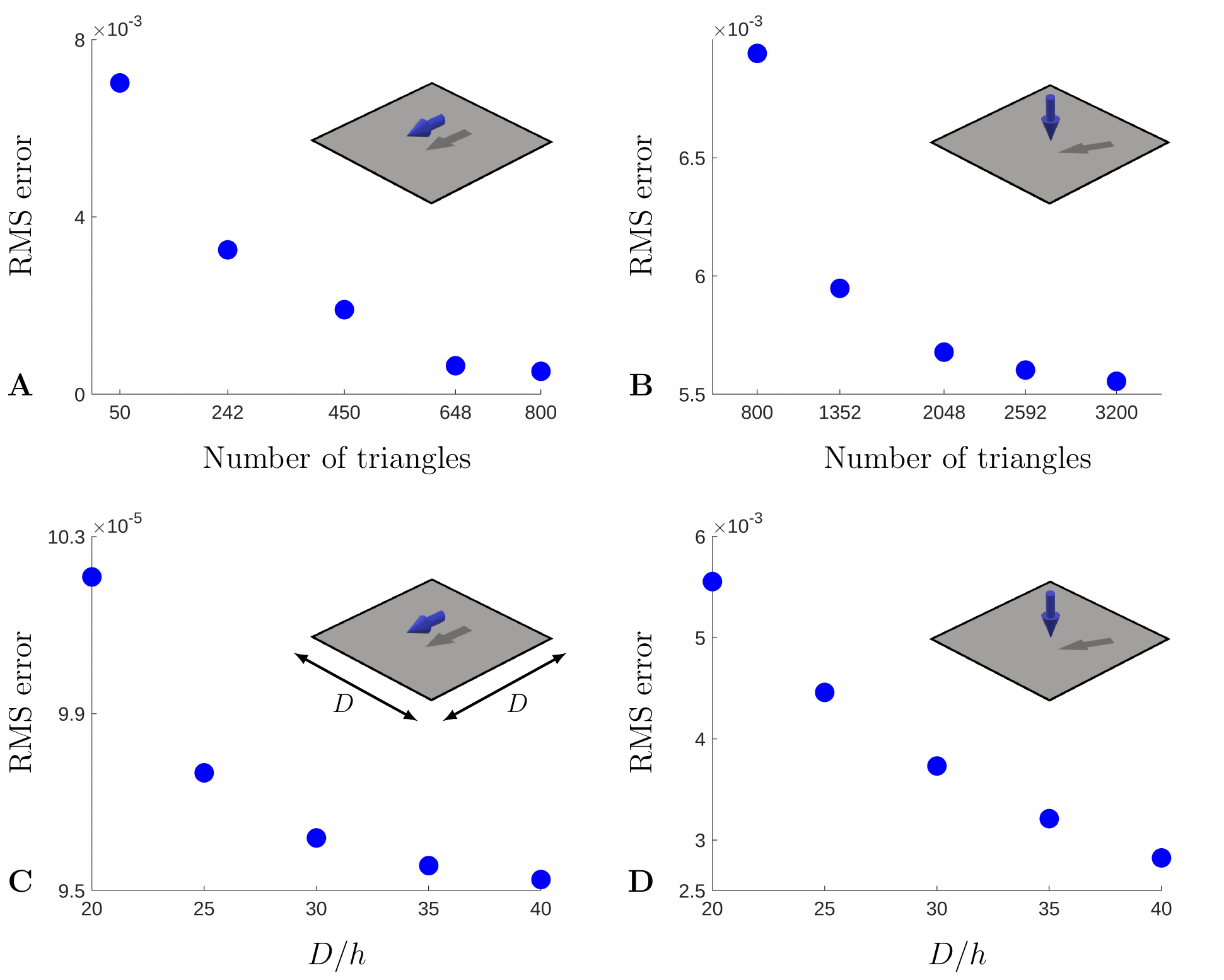}
			\caption{Validation of the BEM code on the case of a point force near a plane wall. We show the convergence towards the exact solution by plotting the RMS error in the cases of a point force parallel  (A,C) and perpendicular (B,D) to an infinite no-slip wall. A,B: the size of the plane is fixed to $D=20h$ and the size of the discretising elements is varied. C,D: the size of the elements is fixed but the size of the discretised part of the plane $D$ is varied.}\label{fig::error}
		\end{figure*}	
		
		\subsection{Point force near a plane wall}
		Consider first a three-dimensional  semi-infinite domain of fluid $(x_1,x_2,x_3)$ with $x_3>0$ and  a solid no-slip boundary  located at $x_3=0$. If a point force $\bm{F}$ is set at $\bm{y} = (0,0,h)$ the exact flow can be expressed  in index notation (with the summation convention assumed) as~\cite{blake1971note}
		\begin{equation}
			(u_E)_j = \frac{F_k}{8\pi\mu}\left[-\frac{\delta_{jk}}{r}-\frac{r_j r_k}{r^3} +2h(\delta_{k\alpha}\delta_{\alpha l}-\delta_{k3}\delta_{3l})\frac{\partial}{\partial r_l}\left\{\frac{hr_j}{r^3}-\left(\frac{\delta_{j3}}{r}+\frac{r_jr_3}{r^3}\right)\right\}\right],
		\end{equation}
		with  $j,k,l = 1,2,3$ while $\alpha = 1,2$. The image point in this case is simply located at the  reflection of $\bm{y}$ below the surface,  i.e.~$\bm{y}^* = (0,0,-h)$.
		
		For our numerical computations, we uniformly triangulate a part of the $x_3=0$ plane where $-{D/2}\leq x_1,x_2 \leq {D/2}$ and apply our BEM code to find a numerical approximation of the flow $\u_N$.
		To define our RMS error,  we consider a cubic lattice of points discretising a cube centred at $(0,0,h)$, with edges of length $h$ parallel to the coordinate axes. The dependence of the RMS error {on the size of the finite elements} is plotted in Fig.~\ref{fig::error}A for a point force parallel to a surface and Fig.~\ref{fig::error}B in the 
		perpendicular case. {For these computations we discretise the $D \times D$ part of the plane, with $D = 20h$.} The error is of range $\approx 10^{-2}-10^{-3}$, and  with a decrease of the mesh size (i.e.~an increase of the total number of triangles in a discretisation), we obtain a monotonic decrease in the error for both orientations of the force. {Additionally, we confirm the convergence to the exact solution by keeping the size of the triangles fixed but varying the size of the discretised part of the plane $D$, see Figs.~\ref{fig::error}C,D.}
		
		Note that in the  case of a perpendicular force, we need to discretise the plane much finer than in case of a parallel force to get a monotonic decay of the error. This results from the different spatial decay of the total flow in the two cases; when the force is parallel, the total far-field flow is a force dipole with a $1/r^2$ decay, while in the perpendicular case the far-field is a force quadrupole and a source dipole decaying as $1/r^3$. The single layer potentials, crucial to the BEM scheme \cite{pozrikidis_1992}, follow the decay of the total flow and thus the gradients of the potential in the perpendicular case vary more rapidly than in the parallel case. Hence, the perpendicular orientation of the force requires a finer mesh to capture the rapid variations of the single layer potential.
		
		\begin{figure*}[t]
			\includegraphics[width=\linewidth]{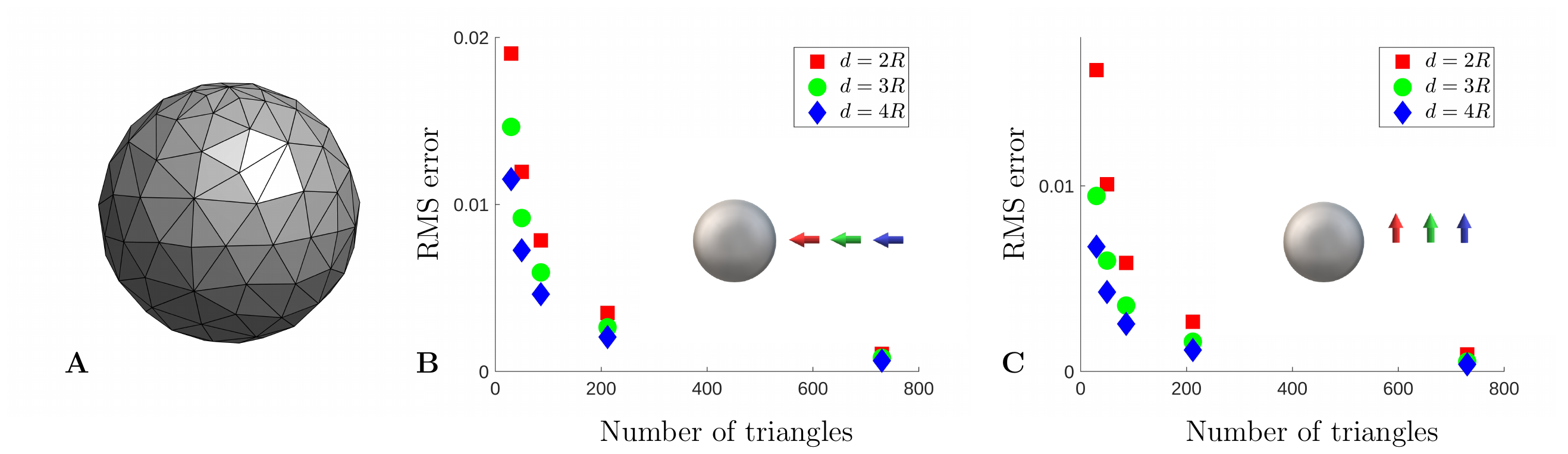}
			\caption{Validation of the BEM code on the case of a point force near a rigid sphere. A: An example of the discretisation of the sphere. B,C:   RMS error as a function of the number of triangles used to discretise the sphere, in the case of a point force being perpendicular (B) or parallel (C) to the surface of the sphere.}\label{fig::error_sph}
		\end{figure*}	
		
		\subsection{Point force near a sphere}
		
		Consider now an infinite fluid domain $|\bm{x}|>R$ outside  a rigid no-slip   sphere located at $|\bm{x}| = R$. 	 The  point force $\bm{F}$ is located at $\bm{y}=(d,0,0) = d\hat{\bm{d}}$, where $d>R$. In  the spherical geometry, the image point is obtained by the inversion of $\bm{y}$ with respect to the sphere i.e. $\bm{y}^* = (R^2/d,0,0)$. {In the case of a radial force, i.e.~one with $\bm{F} = (F,0,0) = F\hat{\bm{d}}$,} the exact solution for the image flow is given by~\cite{KimKarrila} 
		\begin{eqnarray}
			\nonumber {\bf u}_E = -\left(\frac{3R}{2d}-\frac{R^3}{2d^3}\right)F \hat{\bm{d}} \cdot \bm{G}(\bm{r})-F\left[\left(\frac{R^2}{d^2}-\frac{R^4}{d^4}\right)\left(\hat{\bm{d}}\hat{\bm{d}}-\frac{1}{3}\mathbb{I}\right)\cdot\bm{\nabla}\right]\cdot\bm{G}(\bm{r}) \\ -\frac{FR}{4d}\left(1-\frac{R^2}{d^2}\right)^2 \hat{\bm{d}} \cdot \nabla^2 \bm{G}(\bm{r}).
		\end{eqnarray}
		{The flow in the case of a transverse   force has a closed form as well \cite{Oseen,SphereStokeslet} but it is cumbersome and hence we do not include it here.
			
			The sphere is meshed using the FreeFEM++ software \cite{freefem} by discretising the domain of   standard spherical coordinates, adjusting the mesh to the metric of the sphere and then mapping it to its  surface. This process produces relatively uniform elements, as illustrated in Fig.~\ref{fig::error_sph}A. We define the RMS error with the exact solution using a uniformly spaced set of points on the surface of the sphere $|\bm{x}| = 5R$. In Figs.~\ref{fig::error}B,C we plot the value of this error for computations with three different values of the separation between the point force and the sphere ($d/R=2,3,4$), in the case of a radial and transverse force, respectively}. The error is on the order of $10^{-2}$ and is seen to decay	monotonically  with an increase in the number of triangles  uses to discretise the spherical boundary $|\bm{x}|=R$. 	Note that the errors decrease as the force is further away from the spherical boundary. This is to be expected since the RMS error is absolute and the image flow becomes weaker in magnitude with an increase of $d/R$.
		{
			\subsection{Code robustness to the geometry of the cross-section}
			
			The numerical results shown in Fig.~\ref{fig::main_fig} were obtained by using a mesh with 5 vertices on each discretised cross-section. We also 
			performed the same computations for  meshes  with 3 or with 4 vertices on each cross-section. In Fig.~\ref{fig::error_lvar} we show the relative error between the computational results for $d \ln |\bm{u}_S^{(1)}| / d \lambda$ obtained with a pentagonal mesh and the other two types of meshes (triangle and square); the error is well below a few percent for the range of separations  relevant for our study. 
			\begin{figure*}[t]
				\includegraphics[width=\linewidth]{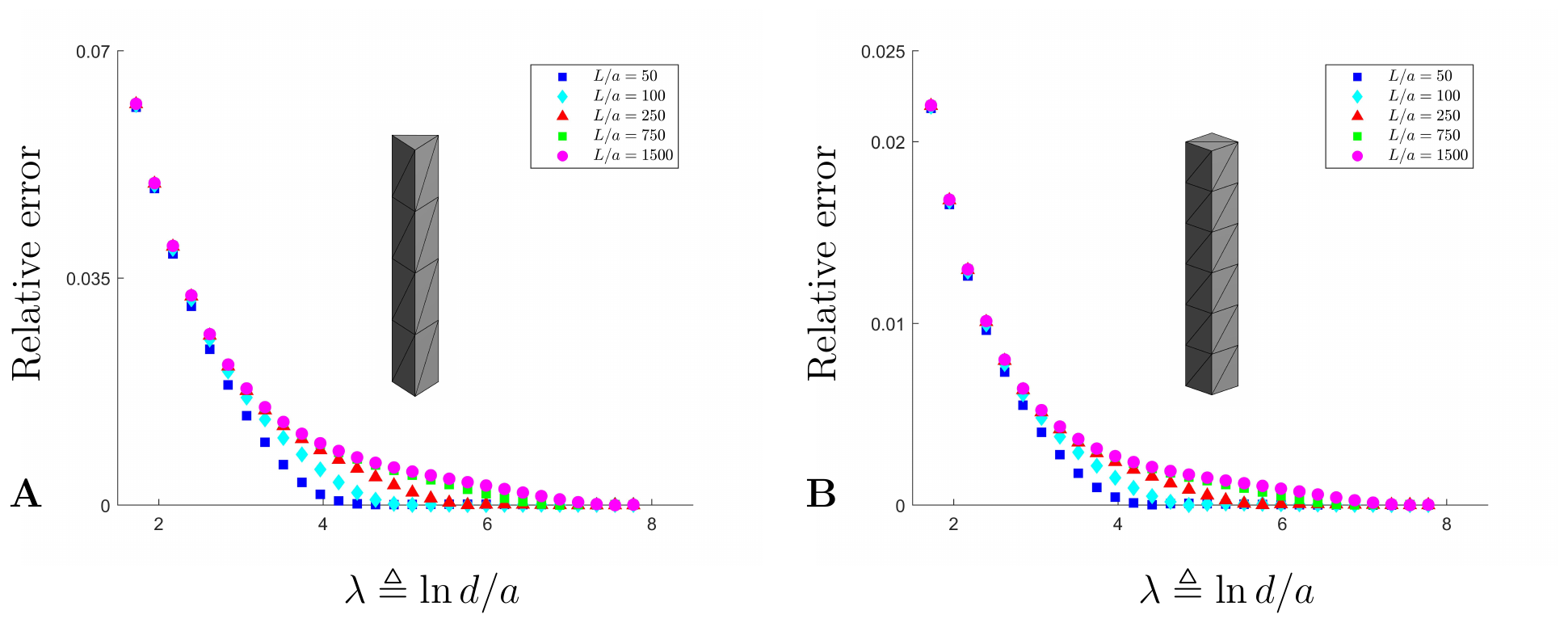}
				\caption{Relative error in computations of $d \ln |\bm{u}_S^{(1)}| / d \lambda$ between results obtained with a mesh of pentagonal cross-sections and meshes with triangular (A) and square (B) cross-sections.}\label{fig::error_lvar}
			\end{figure*}	
		}
	}

\end{document}